\documentclass[%
 reprint,
 amsmath,amssymb,
prb,
]{revtex4-2}

\usepackage{graphicx}
\usepackage{dcolumn}
\usepackage{bm}

\begin{document}

\title{Normalization procedure for obtaining the local density of states from high-bias scanning tunneling spectroscopy}

\author{Rasa Rejali}
\author{La\"{e}titia Farinacci}
\author{Sander Otte}

\affiliation{Department of Quantum Nanoscience, Kavli Institute of Nanoscience, Delft University of Technology, 2628 CJ Delft, The Netherlands
}

\begin{abstract}
Differential conductance spectroscopy performed in the high bias regime---in which the applied voltage exceeds the sample work function---is a poor measure of the local density of states due to the effects of the changing tunnel barrier. Additionally, the large applied voltage oftentimes makes constant-height measurement experimentally impractical, lending constant-current spectroscopy an advantageous edge; but the differential conductance in that case is even further removed from the local density of states due to the changing tip height. Here, we present a normalization scheme for extracting the local density of states from high bias scanning tunneling spectroscopy, obtained in either constant-current or constant-height mode. We extend this model to account for the effects of the in-plane momentum of the probed states to the overall current. We demonstrate the validity of the proposed scheme by applying it to laterally confined field-emission resonances, which appear as peak-shaped spectroscopic features with a well-defined in-plane momentum.  
\end{abstract}

\maketitle

\section{Introduction} 

Scanning tunneling spectroscopy is a tool for obtaining atomically resolved information about the surface-projected electronic density of states as a function of energy. However, the measured differential conductance depends on a number of parameters---such as the transmission through the tunnel barrier, the tip density of states, the tip-sample distance, and the finite temperature---which complicates its direct translation to a local density of states (LDOS). Several schemes have been proposed for extracting quantitative information from the tunnel current, for instance by considering the static conductance $\mathrm{d} \ln I /\mathrm{d} \ln V$~\cite{StroscioPRL1986}, fitting the differential conductivity to a tunneling probability function~\cite{UkraintsevPRB1996}, or normalizing the differential conductivity by both the tunnel current and the transmission coefficient~\cite{KoslowskiPRB2007,PassoniPRB2009}. However, these approaches are solely suited for the treatment of spectroscopic data obtained at a constant tip-sample distance, for electron energies well below the sample and tip work-functions. 

In general, constant height measurements are a better approximation of the local density of states~\cite{ZieglerPRB2009,HormandingerPRL1994,KrennerSciReports2013}, but performing constant-current spectroscopy can be an advantageous choice when, for example, covering large voltage ranges, or when the change in the apparent height of the surface varies drastically---both scenarios necessitate a large dynamic range of current if the tip-sample distance is held constant. Also, high currents during spectroscopy can sometimes alter or damage molecules or the atomic structure at the surface, as well as induce tip instabilities, in which case it can be preferable to maintain a low and constant current. Unfortunately, constant-current spectra are quite removed from the LDOS, in part due to the changing transmission through the tunnel barrier, as well as the effects of the tip displacement during data acquisition: for instance, the energy, relative amplitude, and spatial extent of spectroscopic features are significantly affected by changes in tip height~\cite{ZieglerPRB2009,KrennerSciReports2013}. As such, it is critical to normalize constant-current spectra to reliably extract quantitative information from the measured differential conductance. While such schemes have been proposed~\cite{ZieglerPRB2009,KrennerSciReports2013,HellenthalPRB2013,PronschinskePRB2011}, they are limited to low voltage ranges, well below the sample work-function---even though constant-current measurements are most severely needed in the high bias regime. 

A final point of consideration is the $k$-selectivity of the tunnelling electrons: in a scanning probe configuration, the tunneling current is exponentially dependent on the tip-sample distance, as well as the in-plane momentum of the probed states~\cite{TersoffPRL1983,TersoffPRB1985}, making differential conductance measurements mostly sensitive to states with small in-plane momentum~\cite{KraneSurfSci2018}. 

Here, we present a normalization scheme for extracting the local density of states from spectroscopic data obtained at either constant current or constant tip-sample distance in the field-emission regime, where the applied bias voltage exceeds the sample work function. We extend this model to account for the $k$-selectivity of tunneling electrons, and apply it to spectroscopic measurements---performed in both measurement modes---of confined field emission resonances. The particle-in-a-box modes generated by this confinement have a well-defined in-plane momentum~\cite{Rejali2022}, a characteristic that makes them ideally suited to normalization via a $k$-dependent scheme.

\section{Model}
\label{section:model}
To determine the relation between the measured differential conductance and the local density of states, we begin by describing the total tunneling current $I$ using the one-dimensional Wenzel-Kramers-Brillouin (WKB) theory, wherein the tunneling barrier is defined by a transmission coefficient, $\mathcal{T}(z, V, E)$, that is both energy- and distance-dependent. In the low temperature limit (where the temperature $T\ll eV/k_B$, $e$ is the electron charge and $k_B$ the Boltzmann constant), applying a bias voltage $V$ across the barrier leads to a tunneling current at a tip-sample distance $z$ that is determined by~\cite{FeuchtwangPLA1983, UkraintsevPRB1996,SimmonsJAP1963}: 
\begin{equation}
    I(z, V) = \int_{0}^{eV} \rho_s(E)\rho_t(E-eV)\mathcal{T}(z, V, E) dE, 
    \label{eq:current}
    \end{equation}
where we set the proportionality constant relating the current to the integral to unity, and $\rho_s$ and $\rho_t$ are the tip and sample densities of states, respectively. 

To determine the expression for the differential conductance performed at a constant tip-sample distance $z_0$, we simplify the tip density of states by approximating it to be constant~\cite{HellenthalPRB2013} and setting it equal to unity~\cite{KoslowskiPRB2007}, arriving at:  
\begin{align}
\begin{split}
\mathrm{d}_V I (z_0,V)= & e\rho_s(eV)\mathcal{T}(z_0, V, eV) \\
    & + \int_{0}^{eV} \rho_s(E)  \partial_V \mathcal{T}(z_0, V, E) dE,
    \label{eq:dIdV_halfway}
\end{split}
\end{align}
where we apply the Leibniz rule to differentiate the argument of the integral.
Analogously, we can express the differential conductance obtained in constant current mode as:
\begin{align}
    \begin{split}
    \mathrm{d}_V I (z(V), V)  &= e\rho_s(eV)\mathcal{T}(z(V), V, eV) \\ 
    & + \int_{0}^{eV} \rho_s(E)  \Bigl[\partial_V \mathcal{T}(z(V), V, E) \\
    & \hspace{1.5cm} + \mathrm{d}_V z(V) \partial_z \mathcal{T}(z(V), V, E)\Bigr] dE,
    \end{split}
    \label{eq:dIdV_start_CC}
\end{align}
where we have accounted for voltage dependence of the transmission through $z(V)$ explicitly:
\begin{align}
\begin{split}
    \mathrm{d}_V\mathcal{T}(z(V), V, E) &= \partial_V \mathcal{T}(z(V), V, E) \\
    & + \mathrm{d}_V z(V) \partial_z \mathcal{T}(z(V), V, E).
\end{split}
\end{align}
We note that both expressions for the differential conductance (Eqs.~\ref{eq:dIdV_halfway} and ~\ref{eq:dIdV_start_CC}) involve the derivative of the transmission with respect to the applied bias voltage: these terms cannot be neglected in the high bias regime. 

\begin{figure}[!ht]
    \centering
    \includegraphics{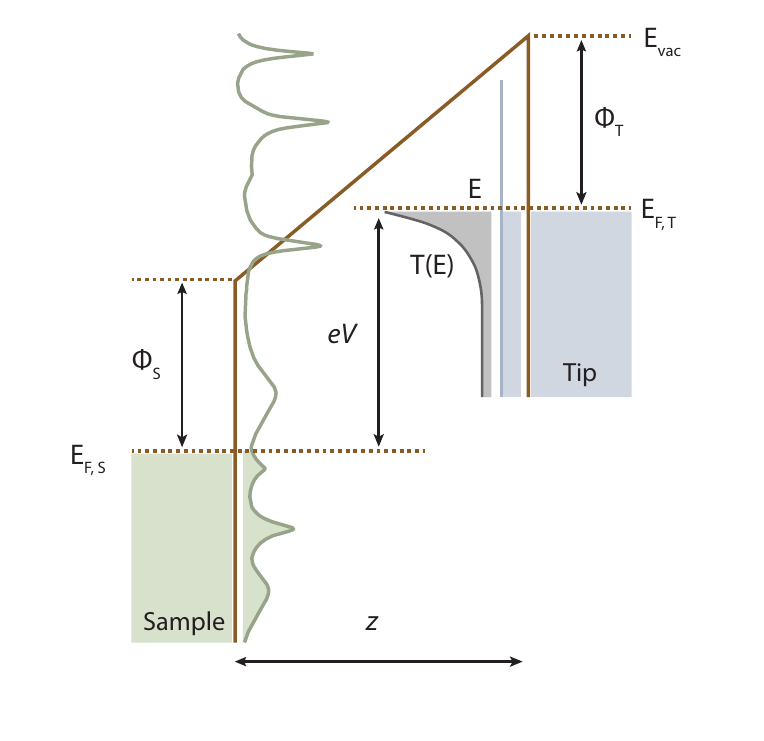}
    \caption{\textbf{Energy landscape at the tip-sample junction, in the direction perpendicular to the sample surface}. \textbf{a} A positive bias voltage applied to the sample shifts the sample Fermi level ($E_{F,S}$) by $-eV$ relative to the tip Fermi level ($E_{F,T}$), creating a trapezoidal potential barrier. This barrier, which depends on the tip-sample distance $z$, and the work functions of the tip ($\phi_t$) and sample ($\phi_s$), has an associated transmission factor $T(E)$, which indicates the exponentially decreasing probability that an electron with a certain energy tunnels through the barrier. The Fermi-Dirac distributions of the sample and tip (green and blue rectangles, respectively) are schematically shown for zero temperature. Electrons from the occupied states of the tip tunnel to the empty states on the sample side. The tip density of states (blue line) is assumed to be a constant, while the sample density of states can be expected to vary with energy (schematically illustrated by the green line); both are filled below the respective Fermi levels. }
    \label{fig:energy_normalization}
\end{figure}

To further evaluate Eqs.~\ref{eq:dIdV_halfway} and ~\ref{eq:dIdV_start_CC}, we need an analytical form for the transmission factor in the field emission regime. In this case, the applied voltage is by definition greater than the sample work function, and we can approximate the tunnel barrier to be triangular, in the region where $eV>\phi_s$, with a transmission factor given by the WKB approximation, wherein the transmission factor $\mathcal{T}$ is related to the integral of the momentum $p$ via $\mathcal{T}(z, V, E) = \exp\left({\frac{-2}{\hbar} \int |p(z')| dz}\right)$:
\begin{equation}
\begin{split}
    -\frac{2}{\hbar}& \int_{z_{\rm{tp}}}^{z}|{p(z')| dz'} = \\  
    & -\frac{2}{\hbar}\int_{z_{\rm{tp}}}^{z}\sqrt{2m_e\left( \phi_s + \frac{\phi_t- \phi_s+ eV}{z}z' - E\right)}dz',\\
\end{split}
\end{equation}
where the classical turning point is denoted by $z_{\rm{tp}}$ and the electron mass by $m_e$~\cite{LandauLifshitz}. The width of the barrier is given by the tip-sample distance $z$, and $\phi_s$ and $\phi_t$ are the tip and sample work functions, respectively (see Fig.~\ref{fig:energy_normalization}). 
From this we can derive a transmission~\cite{NordheimFowler1928,LandauLifshitz}:
\begin{equation}
\begin{split}
     \mathcal{T}(z, V, & E)  = \\
     & \exp \left( -\frac{4\sqrt{2m_e}}{3\hbar}\frac{z}{\phi_t- \phi_s+ eV} (\phi_t + eV - E)^{3/2} \right).
     \label{eq:transmission}
\end{split}
\end{equation}
Here, we note a key and convenient relation~\cite{KoslowskiPRB2007} between the transmission factor $\mathcal{T}(z, V, E)$ through the tunnel barrier, and its partial derivative with respect to the applied voltage $\partial_V\mathcal{T}(z, V, E)$:
\begin{align}
\begin{split}
     \partial_V \mathcal{T}(z, V, & E)  =  \frac{2e\sqrt{2m_e}}{\hbar}\frac{z}{(\phi_t - \phi_s + eV)} \mathcal{T}(z, V, E) \\
    & \times \left(
    \frac{2(\phi_t + eV - E)^{3/2}}{3(\phi_t - \phi_s + eV)} - (\phi_t + eV - E)^{1/2}\right).
    \end{split}
\end{align}
A similar relation can be obtained between the transmission factor $\mathcal{T}(z, V, E)$ and its partial derivative with respect to the tip-sample distance $\partial_z\mathcal{T}(z, V, E)$:

\begin{align}
\begin{split}
   \partial_z \mathcal{T}(z(V), &  V, E)=   \\
   & - \frac{4\sqrt{2m_e}}{3\hbar}\frac{(\phi_t + eV - E)^{3/2}}{\phi_t - \phi_s + eV} \mathcal{T}(z(V), V, E). 
\end{split}
\end{align}

We can now substitute these expressions in Eq.~\ref{eq:dIdV_halfway} and Eq.~\ref{eq:dIdV_start_CC} to determine the relationship between the local density of states and the differential conductance obtained in either measurement mode. 

First, we consider differential conductance spectroscopy performed in constant height mode (Eq.~\ref{eq:dIdV_halfway}), for which we obtain:
\begin{align}
    \begin{split}
    \mathrm{d}_V I(z_0, & V)  = e\rho_s(eV)\mathcal{T}(z_0, V, eV) \\
    & +\int_{0}^{eV} \rho_s(E)  \mathcal{T}(z_0, V, E) \\
    &  \times \frac{2e\sqrt{2m_e}}{\hbar}\frac{z_0}{(\phi_t - \phi_s + eV)} \\
    & \times \left(
    \frac{2(\phi_t + eV - E)^{3/2}}{3(\phi_t - \phi_s + eV)} - (\phi_t + eV - E)^{1/2}
    \right) dE.
    \end{split}
    \label{eq:didv_comp}  
\end{align}

The arguments of the integrals in Eq.~\ref{eq:current} and Eq~\ref{eq:didv_comp} are the same (noting that we set the tip density of states to unity), except for the additional factors that relate the transmission to its partial derivative. By noting that these factors are slowly varying in the energy range of interest, and the transmission itself exponentially peaks at an energy $eV$, we can set their mean value by evaluating them at $E=eV$, yielding $\frac{4e\sqrt{2m_e}}{3\hbar}\frac{z}{(\phi_t - \phi_s + eV)^2}  \phi_t^{3/2}$ and $-\frac{2e\sqrt{2m_e}}{\hbar}\frac{z}{\phi_t - \phi_s + eV} \phi_t^{1/2}$, respectively. Using the generalized mean value theorem~\cite{KoslowskiPRB2007} to evaluate the integral, we obtain:
\begin{align}
\begin{split}
    \mathrm{d}_V I(z_0, V) =  & e\rho_s(eV)\mathcal{T}(z_0, V, eV) + \\ & \frac{4\sqrt{2m_e}}{3\hbar}\frac{z_0}{\phi_t - \phi_s + eV} \\
    & \times \left( \frac{\phi_t^{3/2}}{\phi_t - \phi_s + eV} -\frac{3}{2} \phi_t^{1/2} \right) e I(z_0, V).
\end{split}
\label{eq:didv_general}
\end{align}
Using this expression, we can isolate the density of states from a constant-height differential conductance measurement according to:
\begin{equation}
    \begin{split}
    \rho_s( & eV ) =  \frac{1}{e\mathcal{T}(z_0, V, eV)}\Biggl(\mathrm{d}_V I(z_0, V)  \\
    &  - \frac{4\sqrt{2m_e}}{3\hbar}\frac{z_0 e I(z_0, V)}{\phi_t - \phi_s + eV} \left( \frac{\phi_t^{3/2}}{\phi_t - \phi_s + eV} -\frac{3}{2} \phi_t^{1/2} \right) \Biggr).
    \end{split}
     \label{eq:rho_ch}
\end{equation}

Equivalently, we can evaluate the differential conductance for constant current measurements to isolate the local density of states. Noting once more that the factors that relate the transmission to its derivative with respect to $z(V)$ are slowly varying, we can again apply the mean value theorem to evaluate the integral:
\begin{align}
    \begin{split}
        \mathrm{d}_V  I  (z(V), & V)  =   e\rho_s(eV)\mathcal{T}(z(V), V, eV)\\
        & +\frac{4\sqrt{2m_e}}{3\hbar} \frac{\phi_t^{3/2} I_0}{\phi_t - \phi_s + eV}\\
        &   \times \left(
        ez(V) \left(\frac{1}{\phi_t - \phi_s + eV} - \frac{3}{2\phi_t }\right) - \mathrm{d}_V z(V)\right), 
    \end{split}
\end{align}
where $I_0$ is the current set-point. Here, we should note that a measurement of the $\mathrm{d}I/\mathrm{d}V$ at a specific dc-bias amounts to tracking the change in the current caused by the applied bias modulation; correspondingly, the $\mathrm{d}_V z(V)$ term corresponds  to the change in the tip-sample separation due to the same ac component of the applied bias. If the frequency of the ac-bias is sufficiently high compared to the cut-off frequency of the feedback, which is normally the case, then $\mathrm{d}_V z(V)$ is negligible~\cite{ZieglerPRB2009,HellenthalPRB2013}. In this case, the density of states can be extracted from the measured differential conductance via:

\begin{equation}
    \begin{split}
        \rho_s(eV) &= \frac{1}{e\mathcal{T}(z(V), V, eV)} \Biggl(\mathrm{d}_V I(z(V), V) \\ & - \frac{4\sqrt{2m_e}}{3\hbar}\frac{z(V)e I_0}{\phi_t - \phi_s + eV} \left( \frac{\phi_t^{3/2}}{\phi_t - \phi_s + eV} -\frac{3}{2} \phi_t^{1/2} \right) \Biggr)   
    \end{split}
    \label{eq:rho_cc}
\end{equation}

The relations in Eq.~\ref{eq:rho_ch} and \ref{eq:rho_cc} can be used to normalize constant-height and constant-current differential conductance spectra, respectively, to obtain the local density of states, in the case where the applied voltage is greater than the sample work-function. This approach relies on recording the differential conductance $\mathrm{d}_V I(z, V)$, the relative tip-displacement $\Delta z$, and the current-voltage behaviour $I(V)$ simultaneously. While this is easily and commonly implemented, gaining experimental information about the absolute distance $z = \Delta z + z_0$ requires additional $I(z)$ measurements to estimate the point of contact between the tip and sample. 

\begin{figure*}[!ht]
    \centering
    \includegraphics[width = 0.75\textwidth]{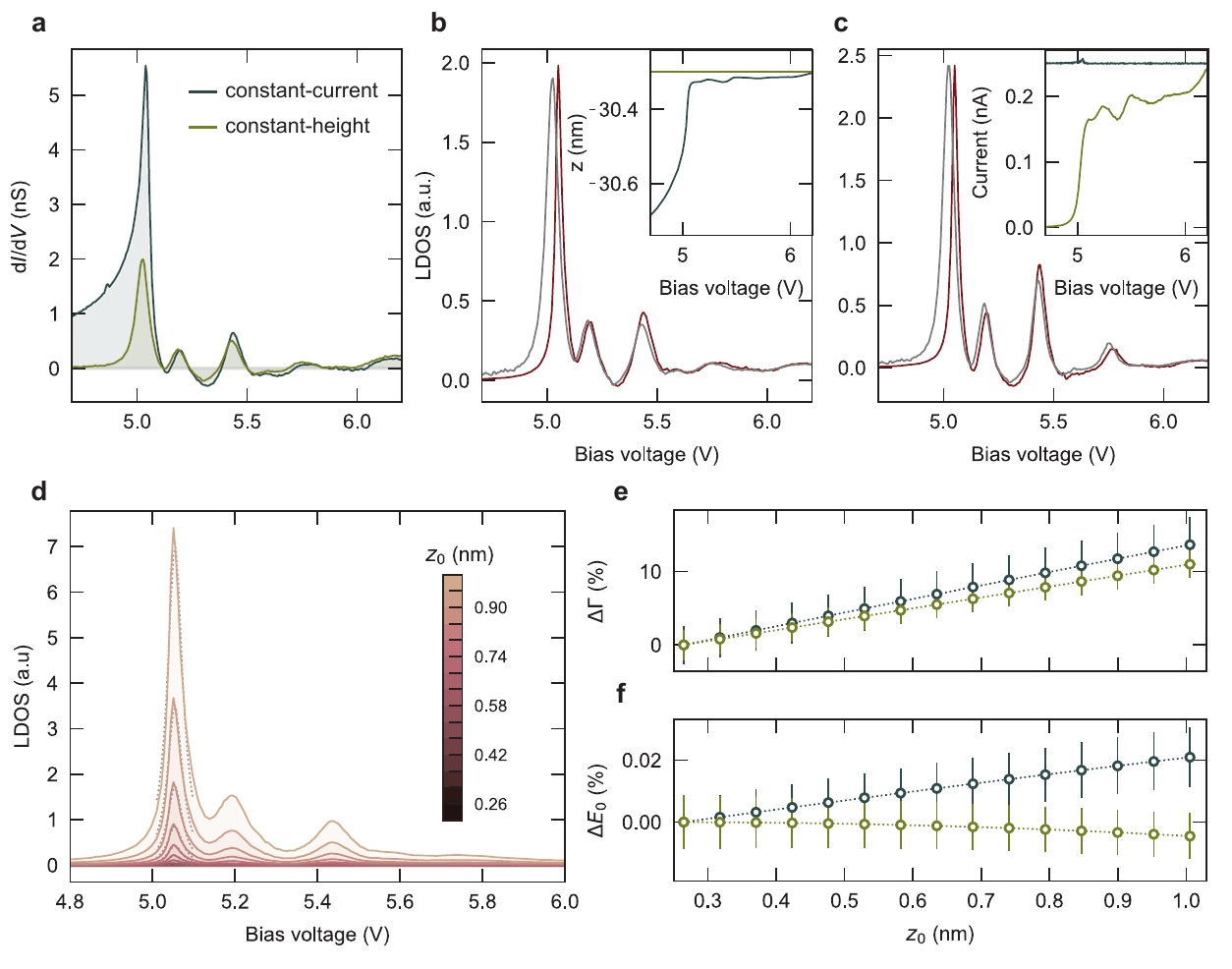}
    \caption{\textbf{Local density of states extracted from constant-height and constant-current spectroscopic data}. \textbf{a} Constant-height (light green) and constant-current (dark blue) differential conductance spectroscopy probing confined field-emission resonances on the chlorinated Cu(100) surface~\cite{Rejali2022}, obtained at current set-point of 250pA. \textbf{b, c} The raw d$I$/d$V$ is normalized to obtain the local density of states, with (c) and without (b) considering the in-plane momentum of the probed state, for both the constant-height (light gray) and constant-current (maroon) measurements. Simultaneously acquired tip-sample distance, offset by $z_0$ to obtain the absolute distance (b, inset), and current-voltage curve (c, inset), measured during both constant-height (light green, $z_0 \approx 0.2$~nm) and constant-current (grey, $z_0 \approx 0.6$~nm) spectroscopy. \textbf{d} Local density of states obtained from normalizing the constant-height spectrum, for a range of $z_0$ (exact values indicated in the color-bar). Dotted lines are Lorentzian profiles fitted to the first peak. \textbf{e, f} Effects of a changing $z_0$ on the relative peak width (e) and energy (f) of the first peak in the LDOS, extracted from both constant-current (dark blue) and constant-height (light green) spectra.}
    \label{fig:diffNorm}
\end{figure*}

However, it is in general possible to make a reasonable estimate of this parameter~\cite{ZieglerPRB2009}. In the particular case of field-emission resonances, the absolute distance can be estimated by modelling the out-of-plane potential to match the experimental and calculated energies of the resonances~\cite{Rejali2022}. In either case, the exact value of $z_0$ does not dramatically affect the spectral shape of the normalized differential conductance: for a wide range of $z_0$ values, the peak positions and widths remain roughly constant, while the relative height of the peaks are subject to variation (see Fig~\ref{fig:diffNorm})~\cite{ZieglerPRB2009}. This demonstrates the robustness of the normalization scheme against uncertainties in $z_0$.

\section{Effects of $k$-selectivity in the tunneling current}

In  general, the total tunneling current depends not only on the tunneling barrier, but also on the in-plane momentum $k_{\parallel}$ of the probed state~\cite{TersoffPRL1983,TersoffPRB1985}. In fact, scanning tunneling spectroscopy is mostly sensitive to states with a small in-plane momentum, meaning the total contribution to the current dies off as the $k_{\parallel}$ of the state increases~\cite{KraneSurfSci2018}. This effect has been previously accounted for in the Bardeen description of the tunneling current: there, $k_{\parallel}$ is incorporated into the decay constant that defines the current. Namely, $I \propto \exp({-2\kappa z})$, where $\kappa = \sqrt{2m\phi/\hbar^2 + k_{\parallel}^2}$, and $\phi$ is the potential barrier for tunneling~\cite{KraneSurfSci2018, TersoffPRL1983,TersoffPRB1985}. 

Analogously, the effects of the in-plane momentum can be accounted for in the WKB approach via the transmission factor, namely:
\begin{equation}
    \begin{split}
         \mathcal{T}(z, V,& E, k_{\parallel}) \\
         &= \exp\left({\frac{-2}{\hbar} \int_0^z \sqrt{2m_e(\phi(z') - E) + \hbar^2 k_{\parallel}^2 }dz'}\right).
    \end{split}
\end{equation}
In the simplest case, where the tunneling barrier is approximated by a rectangular potential ($eV < \phi_s, \phi_t$, necessarily), with an effective tunneling barrier height $\phi_{\rm{eff}} = (\phi_t + \phi_s)/2$, the transmission reduces to $\mathcal{T}(z, V, E) = \exp\left({\frac{-2}{\hbar}z \sqrt{2m_e(\phi_{\rm{eff}}- E) + \hbar^2 k_{\parallel}^2 }}\right)$. The tunneling current in this case is proportional to this transmission:
\begin{align}
    I(V) & \propto \mathcal{T}(z, V, E=E_F, k_{\parallel}) \int_0^{eV} \rho_s(E) dE, \\
    & \propto \exp\left({\frac{-2}{\hbar}z \sqrt{2m_e\phi_{\rm{eff}} + \hbar^2 k_{\parallel}^2 }}\right)\int_0^{eV} \rho_s(E) dE,
\end{align}
as in the Bardeen approach. In the above, we consider $\mathcal{T}(z, V, E, k_{\parallel}) \sim \mathcal{T}(z, V, E=E_F, k_{\parallel})$, a common approximation when $eV < \phi_s, \phi_t$, and we set the tip-density of states to unity.

In the field-emission regime, we can similarly incorporate the effects of the in-plane momentum of the probed state, leading to a transmission factor that depends on $k_{\parallel}$:
\begin{equation}
    \begin{split}
    \mathcal{T}(z, V,&  E,  k_{\parallel}) =\exp \Biggl( -\frac{2}{3 m_e\hbar}\frac{z}{\phi_t - \phi_s+ eV} \\
    & \times \Bigl((2m_e(\phi_t + eV - E) + \hbar^2 k_{\parallel}^2)^{3/2} - |\hbar k_{\parallel}|^3\Bigr)\Biggr),
    \end{split}
\end{equation}
where the partial derivative with respect to the voltage is:
\begin{equation}
\begin{split}
    \partial_V  & \mathcal{T}( z, V, E,  k_{\parallel})  =  \mathcal{T}(z, V, E,  k_{\parallel}) \Biggl(\frac{2e}{3 m_e\hbar}\frac{z}{(\phi_t - \phi_s+ eV)^2}\\
    & \times \left(\left(2m_e(\phi_t + eV - E) + \hbar^2 k_{\parallel}^2\right)^{3/2} - \left|\hbar k_{\parallel}\right|^3\right) \\
    &  - \frac{2e}{\hbar}\frac{z}{\phi_t - \phi_s+ eV} \left(2m_e(\phi_t + eV - E) + \hbar^2 k_{\parallel}^2\right)^{1/2}\Biggr).
\end{split}
\end{equation}
From this, we can determine the local density of states by following the same steps as delineated in the previous section to obtain for the density of states from a constant-height measurement:
\begin{align}
    \begin{split}
    & \rho_s(eV) =\frac{1}{\mathcal{T}(z_0, V, eV)}\Biggl( \mathrm{d}_V  I(z_0, V) - \frac{2e}{\hbar}\frac{z_0 I(z_0, V)}{\phi_t - \phi_s+ eV} \\
    &  \times \Biggl( \frac{\left(2m_e\phi_t + \hbar^2 k_{\parallel}^2\right)^{3/2} - \left|\hbar k_{\parallel}\right|^3 }{3m_e (\phi_t - \phi_s+ eV)} - \left(2m_e\phi_t + \hbar^2 k_{\parallel}^2\right)^{1/2}\Biggr) \Biggr),
    \end{split}
    \label{eq:rho_k_ch}
\end{align}
and analogously for constant-current measurements:
\begin{align}
    \begin{split}
    & \rho_s(eV) =\frac{1}{\mathcal{T}(z(V), V, eV)}\Biggl( \mathrm{d}_V  I(z(V),V)- \frac{2e}{\hbar}\frac{z(V) I_0}{\phi_t - \phi_s+ eV} \\
    &  \times \Biggl( \frac{\left(2m_e\phi_t + \hbar^2 k_{\parallel}^2\right)^{3/2} - \left|\hbar k_{\parallel}\right|^3 }{3m_e (\phi_t - \phi_s+ eV)} - \left(2m_e\phi_t + \hbar^2 k_{\parallel}^2\right)^{1/2}\Biggr) \Biggr).
    \end{split}
    \label{eq:rho_k_cc}
\end{align}

\section{Application of normalization procedure to peak-shaped spectroscopic features}

To test the validity of the proposed normalization schemes, we apply them to constant-current and constant-height spectroscopic data obtained for field-emission resonances (FERs). These resonances are quantized electronic states localized in the vacuum, between the surface and the probe tip: the linear potential drop across the junction due to the applied bias voltage elevates the potential barrier above the vacuum level of the sample, thereby giving rise to a new class of confined states. In this high electric field regime, tunneling electrons will be reflected both by the sample surface and the rising potential barrier generated by the applied voltage, thus creating standing waves in front of the sample surface. These vacuum-localized states depend critically on the electronic properties of the sample---such as the surface-projected band structure, which alters the surface reflectivity. As such, they are useful in obtaining information about the surface, including local work function changes~\cite{RuffieuxPRL2009,JungPRL1995,PivettaPRB2005, PloigtPRB2007} or scattering properties at interfaces~\cite{KubbyPRL1990}, and even allow for atomically resolved images of insulators~\cite{BobrovNature2001} and spin-textures~\cite{SchlenhoffAPL2020} far from the surface.

Here, we focus specifically on laterally confined field-emission resonances~\cite{Rejali2022}: we arrange single vacancies on the chlorinated Cu(100) surface~\cite{KalffNatureNano2016,GirovskySciPost2017,DrostNaturePhys2017} to reveal square patches of the underlying metal surface, creating atomically precise potential wells, which we refer to by their size in unit cells. In doing so, we generate particle-in-a-box states that carry some finite $k_{\parallel}$. We can understand this in analogy to the simple case of an infinite potential well, wherein the angular wave number for each state, described by principle quantum number $n$, is given by $n\pi/L$, where $n$ is a positive integer and $L$ is the width of the well. This allows us to test our normalization schemes for a system in which $k_{\parallel}$ is relevant, in addition to inspecting how it compares for open- or closed-feedback measurements.

\subsection{Comparison LDOS extracted from constant-height and constant-current spectroscopy}

Ideally, differential conductance measurements performed in either spectroscopic mode---constant-current or constant-height---should yield, via the proposed normalization procedures, identical local densities of states. In Fig~\ref{fig:diffNorm}, we show d$I$/d$V$ measurements performed in both modes, and normalize each according to Eqs.~\ref{eq:rho_ch}, \ref{eq:rho_cc}, \ref{eq:rho_k_cc} and \ref{eq:rho_k_ch} to obtain the local density of states. The extracted LDOS is remarkably similar for the two measurement modes, for both of the normalization schemes proposed here---which either neglect (Fig~\ref{fig:diffNorm}b, Eqs.~\ref{eq:rho_ch} and \ref{eq:rho_cc}) or consider (Fig~\ref{fig:diffNorm}c, Eqs.~\ref{eq:rho_k_cc} and \ref{eq:rho_k_ch}) the effects of $k$-selectivity on the tunnel current. In both cases, the position, relative amplitude, and width of the peaks in the extracted LDOS are quite consistent, although the first peak appears sharper and slightly shifted for the constant-current normalized spectrum; this is likely due to the drastic change in the tip-sample distance around this resonance, a feedback response that cannot be completely rectified by the normalization scheme.

Explicitly considering the in-plane momentum of the probed states $k_{\parallel}$ (Fig~\ref{fig:diffNorm}c) only affects the relative amplitudes of the resonances, as expected. Since the tunneling current is increasingly less sensitive to states with increasing $k_{\parallel}$, it follows that the relative amplitude of the higher energy peaks are increasingly underestimated if this effect is not accounted for.  

As previously mentioned, the schemes proposed here rely on the absolute tip-sample distance, while usually, only the relative tip displacement $\Delta z$ is readily available and easily measured. The value of $z_0$, which converts the measured relative change into an absolute distance $z = \Delta z + z_0$, can be reasonably estimated by several means---such as additional $I(z)$ measurements---but is hard to pin-point precisely. As such, any normalization procedure that cannot tolerate small variations in $z_0$ cannot be widely implemented.

\begin{figure}[t]
    \centering
    \includegraphics{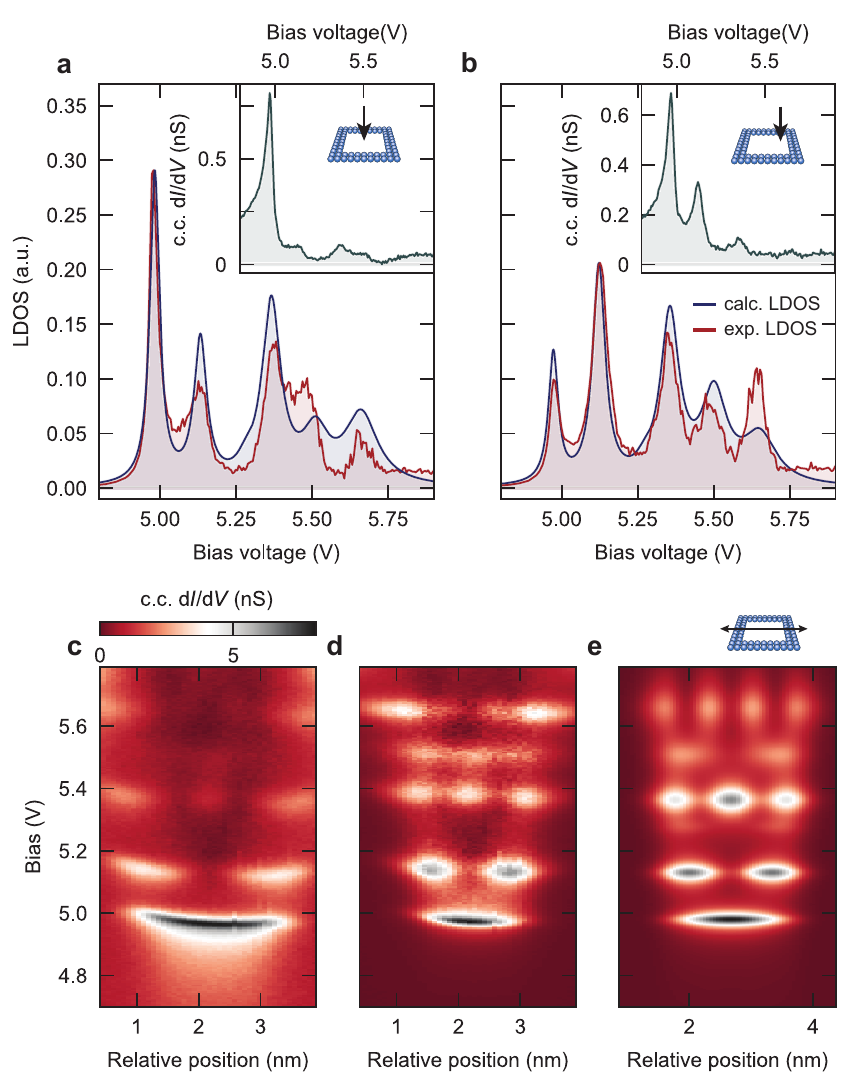}
    \caption{\textbf{Spatially dependent comparison of raw constant-current spectra, and the corresponding extracted and calculated local density of states}. \textbf{a, b} LDOS extracted (red) from constant-current $\mathrm{d}I/\mathrm{d}V$ spectroscopy obtained at a set-point 0f 50~pA  (insets), compared to the calculated LDOS (blue), obtained for the center (\textbf{a}) and edge (\textbf{b}) of the $7\times 7$ patch (the relative position is schematically illustrated, inset). Experimental LDOS is normalized by its maximum value, and re-scaled by the maximum value of the calculated LDOS. \textbf{c} Raw, stacked constant-current differential conductance spectra obtained at a set-point of 50~pA, taken along a line crossing the center of the $7 \times 7$ patch (schematically illustrated to the right). \textbf{d} The corresponding LDOS, extracted from the experimental data using the $k_{\parallel}$ sensitive normalization scheme, and \textbf{e} the calculated LDOS.}
    \label{fig:LDOScomp}
\end{figure}

First, we note that the experimental conditions for each spectroscopic mode help us set some limits on the value of $z_0$ for comparing constant-current and constant-height measurements. Constant-height measurements, for instance, require regulating at a bias voltage of interest $V_{\rm{stpt}}$ (at a current of choice) before opening the feedback for measurement. This is not the case for constant-current measurements, as the feedback is always engaged---but in either case, the tip-sample distance at $V_{\rm{stpt}}$ should be the same for both measurement modes, given measurements are performed with the same microscopic tip and at the same set-point current. Here, we chose $V_{\rm{stpt}}$ to be the highest bias voltage of interest (6.2~V, see Fig~\ref{fig:diffNorm}a), to ensure the tip-sample distance is set to its maximum value over the course of spectroscopy, thereby minimizing the risk of tip crashes. This means that the value of $z_0$, while different for each mode, should result in the same absolute distance $z$ at the highest bias voltage, as we see in Fig~\ref{fig:diffNorm}b (inset). This condition allows us to compare the extracted LDOS for the two measurements, knowing that the parameter $z_0$ does not skew one curve relative to the other.

To more concretely trace the effects of a changing $z_0$, we focus on $\mathrm{d}I/\mathrm{d}V$ spectroscopy normalized for a range of $z_0$ values, as shown in Fig~\ref{fig:diffNorm}d---we can see by eye that this parameter mainly determines the overall intensity of the peaks. To quantify this, we fit the first peak in the experimental LDOS with a Lorentzian lineshape ((Fig~\ref{fig:diffNorm}d) to extract its width and position. Carrying out this procedure for LDOS extracted from both constant-current and -height measurements (Fig~\ref{fig:diffNorm}e, f), we see that in both cases the width, $\Gamma$, and position, $E_0$, of the peak are not significantly altered over a $z_0$ range of nearly $1$~nm---to put this value in context, a few angstroms displacement of the tip can cause the tunneling current to change by an order of magnitude. In fact, we observe a maximum variation of $\Delta E_0\sim 0.02$\% in the peak position ($\sim 5.05$~V at 0.26~nm), which, in this case, is smaller than the broadening associated with the lock-in modulation (10~mV). The change in the peak width $\Delta \Gamma$ is greater, but still negligible over such a broad range of $z_0$: we observe a maximum variation of $\sim 13$\% and $\sim 10$\% in the peak width, for constant-current and constant-height extracted LDOS, respectively. As we can see, the exact value of $z_0$ can impact the overall intensity of the experimentally derived LDOS, but it's the relative change $\Delta z$---which is easily measured during spectroscopy---that plays a critical role in determining the relevant peak features, such as relative intensity, width, and position.

\subsection{Comparison of experimentally and theoretically derived LDOS}

Having ascertained that our normalization procedure is robust for different spectroscopic modes, and yields consistent results against a varying $z_0$, we now focus on how it fares compared to the expected (calculated) local density of states. To calculate the local density of states, we model the potential landscape of the laterally confined field-emission resonances using a finite well with slanted walls where the depth of the well is set by the work-function difference between the chlorinated and bare Cu(100) surfaces and the slope of the walls by the Fermi wavelength~\cite{Rejali2022}. From this, we can calculate the expected eigenstates and energies to determine the LDOS, which is simply given by $|\Psi|^2$. To mimic the state broadening---which is primarily lifetime-limited, but also affected by experimental considerations, such as the lock-in modulation and temperature---we generate Lorentzians centered at each eigenenergy, with the line-width set to match the experiment. More precisely, the lifetime of the resonances decreases as the principle quantum number of the in-plane mode increases~\cite{StepanowPRB2011,CrampinPRL2005_2}, leading to a corresponding increase in the line-width, which is echoed in the theoretical LDOS.

\begin{figure}[ht]
    \centering
    \includegraphics{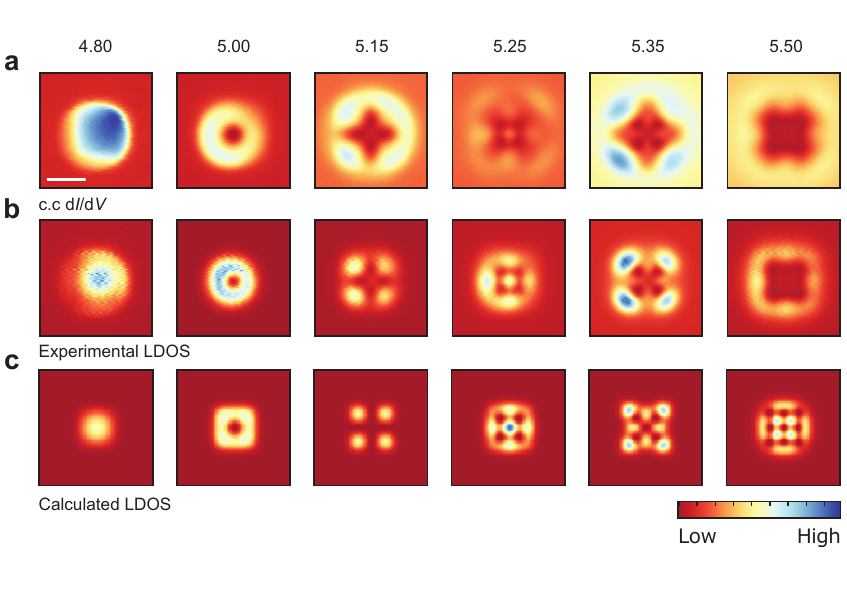}
    \caption{\textbf{Local density of states extracted from constant-current differential conductance maps.} \textbf{a} Experimental constant-current differential conductance maps, taken at a current-set point of 100~pA over the $7\times 7$ patch, at the energies indicated above. Scale bar: 2~nm\textbf. {b} LDOS obtained from normalizing the differential conductance maps compared to \textbf{c} the theoretically derived LDOS.}
    \label{fig:LDOScompmaps}
\end{figure}

Fig.~\ref{fig:LDOScomp} shows the experimental LDOS retrieved from constant-current d$I$/d$V$ measurements performed at the center and edge of square patch made out of $7\times7$ vacancies, compared to the corresponding calculated LDOS. The agreement between the experimentally derived and expected local density of states is remarkable, especially in light of the raw d$I$/d$V$ spectra. The effects of the feedback, which are most pronounced at the energy of the first resonance, are well-distinguished in the raw spectrum obtained at the patch edge (Fig.~\ref{fig:LDOScomp}b): here, the tip displacement artificially lends the first resonance an asymmetric line-shape and heightened relative intensity, which is remedied by the normalization procedure. To probe the spatial evolution of the experimentally derived and calculated LDOS, we perform d$I$/d$V$ spectroscopy along a line crossing the center of the patch (Fig.~\ref{fig:LDOScomp}c); the raw data is normalized to obtain the LDOS (Fig.~\ref{fig:LDOScomp}d), which we find is in fair agreement with the calculated LDOS (Fig.~\ref{fig:LDOScomp}e). This comparison makes it clear that the tip displacement during data acquisition dramatically broadens the spatial extent of the states well beyond the confines of the patch, and changes the line-shape of the spectroscopic features, in agreement with previous findings~\cite{ZieglerPRB2009,HellenthalPRB2013,PronschinskePRB2011}.

Another experimental approach towards probing the full spatial evolution and extent of the local density of states is performing differential conductance maps. To do so in constant-current mode (i.e. with the feedback loop closed) complicates the matter, as the local topographic features will induce significant cross-talk between the tip-sample displacement and the measured differential conductance, convoluting data interpretation~\cite{BerghausSurfSci1988,StroscioKaiser1993,ZieglerPRB2009}. In Fig.~\ref{fig:LDOScompmaps}, we apply our normalization procedure to constant-current maps obtained over a $7\times 7$ patch, and compare it to the theoretically derived LDOS: again, we see a reduction in the spatial extent of the patch, and an increased sharpness in the spectroscopic features that allows us to better distinguish the nodal planes at each energy. 

\section{Conclusions} 

In this work, we present a normalization scheme for extracting quantitative information about the local electronic density of states from constant-current or constant-height spectra taken in the high bias regime, i.e. at bias voltages exceeding the sample work function. We consider the effects of the in-plane momentum of the probed states on this relation, and apply the normalization procedure to laterally confined field-emission resonances. The extracted LDOS obtained through normalization of constant-height and constant-current spectra agree well with each other, and in turn with the theoretically derived LDOS.

\section{Data availability} 
All data presented in this work are publicly available with identifier (DOI) 10.5281/zenodo.6473079

\section{Acknowledgments} 
The authors thank the Dutch Research Council (NWO) and the European Research Council (ERC Starting Grant 676895 “SPINCAD”).

\bibliography{main}

\begin{thebibliography}{31}%
\makeatletter
\providecommand \@ifxundefined [1]{%
 \@ifx{#1\undefined}
}%
\providecommand \@ifnum [1]{%
 \ifnum #1\expandafter \@firstoftwo
 \else \expandafter \@secondoftwo
 \fi
}%
\providecommand \@ifx [1]{%
 \ifx #1\expandafter \@firstoftwo
 \else \expandafter \@secondoftwo
 \fi
}%
\providecommand \natexlab [1]{#1}%
\providecommand \enquote  [1]{``#1''}%
\providecommand \bibnamefont  [1]{#1}%
\providecommand \bibfnamefont [1]{#1}%
\providecommand \citenamefont [1]{#1}%
\providecommand \href@noop [0]{\@secondoftwo}%
\providecommand \href [0]{\begingroup \@sanitize@url \@href}%
\providecommand \@href[1]{\@@startlink{#1}\@@href}%
\providecommand \@@href[1]{\endgroup#1\@@endlink}%
\providecommand \@sanitize@url [0]{\catcode `\\12\catcode `\$12\catcode
  `\&12\catcode `\#12\catcode `\^12\catcode `\_12\catcode `\%12\relax}%
\providecommand \@@startlink[1]{}%
\providecommand \@@endlink[0]{}%
\providecommand \url  [0]{\begingroup\@sanitize@url \@url }%
\providecommand \@url [1]{\endgroup\@href {#1}{\urlprefix }}%
\providecommand \urlprefix  [0]{URL }%
\providecommand \Eprint [0]{\href }%
\providecommand \doibase [0]{https://doi.org/}%
\providecommand \selectlanguage [0]{\@gobble}%
\providecommand \bibinfo  [0]{\@secondoftwo}%
\providecommand \bibfield  [0]{\@secondoftwo}%
\providecommand \translation [1]{[#1]}%
\providecommand \BibitemOpen [0]{}%
\providecommand \bibitemStop [0]{}%
\providecommand \bibitemNoStop [0]{.\EOS\space}%
\providecommand \EOS [0]{\spacefactor3000\relax}%
\providecommand \BibitemShut  [1]{\csname bibitem#1\endcsname}%
\let\auto@bib@innerbib\@empty
\bibitem [{\citenamefont {Stroscio}\ \emph {et~al.}(1986)\citenamefont
  {Stroscio}, \citenamefont {Feenstra},\ and\ \citenamefont
  {Fein}}]{StroscioPRL1986}%
  \BibitemOpen
  \bibfield  {author} {\bibinfo {author} {\bibfnamefont {J.~A.}\ \bibnamefont
  {Stroscio}}, \bibinfo {author} {\bibfnamefont {R.~M.}\ \bibnamefont
  {Feenstra}},\ and\ \bibinfo {author} {\bibfnamefont {A.~P.}\ \bibnamefont
  {Fein}},\ }\bibfield  {title} {\bibinfo {title} {Electronic structure of the
  {S}i(111)2 \ifmmode\times\else\texttimes\fi{} 1 surface by scanning-tunneling
  microscopy},\ }\href {https://doi.org/10.1103/PhysRevLett.57.2579} {\bibfield
   {journal} {\bibinfo  {journal} {Phys. Rev. Lett.}\ }\textbf {\bibinfo
  {volume} {57}},\ \bibinfo {pages} {2579} (\bibinfo {year}
  {1986})}\BibitemShut {NoStop}%
\bibitem [{\citenamefont {Ukraintsev}(1996)}]{UkraintsevPRB1996}%
  \BibitemOpen
  \bibfield  {author} {\bibinfo {author} {\bibfnamefont {V.~A.}\ \bibnamefont
  {Ukraintsev}},\ }\bibfield  {title} {\bibinfo {title} {Data evaluation
  technique for electron-tunneling spectroscopy},\ }\href
  {https://doi.org/10.1103/PhysRevB.53.11176} {\bibfield  {journal} {\bibinfo
  {journal} {Phys. Rev. B}\ }\textbf {\bibinfo {volume} {53}},\ \bibinfo
  {pages} {11176} (\bibinfo {year} {1996})}\BibitemShut {NoStop}%
\bibitem [{\citenamefont {Koslowski}\ \emph {et~al.}(2007)\citenamefont
  {Koslowski}, \citenamefont {Dietrich}, \citenamefont {Tschetschetkin},\ and\
  \citenamefont {Ziemann}}]{KoslowskiPRB2007}%
  \BibitemOpen
  \bibfield  {author} {\bibinfo {author} {\bibfnamefont {B.}~\bibnamefont
  {Koslowski}}, \bibinfo {author} {\bibfnamefont {C.}~\bibnamefont {Dietrich}},
  \bibinfo {author} {\bibfnamefont {A.}~\bibnamefont {Tschetschetkin}},\ and\
  \bibinfo {author} {\bibfnamefont {P.}~\bibnamefont {Ziemann}},\ }\bibfield
  {title} {\bibinfo {title} {Evaluation of scanning tunneling spectroscopy
  data: Approaching a quantitative determination of the electronic density of
  states},\ }\href {https://doi.org/10.1103/PhysRevB.75.035421} {\bibfield
  {journal} {\bibinfo  {journal} {Phys. Rev. B}\ }\textbf {\bibinfo {volume}
  {75}},\ \bibinfo {pages} {035421} (\bibinfo {year} {2007})}\BibitemShut
  {NoStop}%
\bibitem [{\citenamefont {Passoni}\ \emph {et~al.}(2009)\citenamefont
  {Passoni}, \citenamefont {Donati}, \citenamefont {Li~Bassi}, \citenamefont
  {Casari},\ and\ \citenamefont {Bottani}}]{PassoniPRB2009}%
  \BibitemOpen
  \bibfield  {author} {\bibinfo {author} {\bibfnamefont {M.}~\bibnamefont
  {Passoni}}, \bibinfo {author} {\bibfnamefont {F.}~\bibnamefont {Donati}},
  \bibinfo {author} {\bibfnamefont {A.}~\bibnamefont {Li~Bassi}}, \bibinfo
  {author} {\bibfnamefont {C.~S.}\ \bibnamefont {Casari}},\ and\ \bibinfo
  {author} {\bibfnamefont {C.~E.}\ \bibnamefont {Bottani}},\ }\bibfield
  {title} {\bibinfo {title} {Recovery of local density of states using scanning
  tunneling spectroscopy},\ }\href {https://doi.org/10.1103/PhysRevB.79.045404}
  {\bibfield  {journal} {\bibinfo  {journal} {Phys. Rev. B}\ }\textbf {\bibinfo
  {volume} {79}},\ \bibinfo {pages} {045404} (\bibinfo {year}
  {2009})}\BibitemShut {NoStop}%
\bibitem [{\citenamefont {Ziegler}\ \emph {et~al.}(2009)\citenamefont
  {Ziegler}, \citenamefont {N\'eel}, \citenamefont {Sperl}, \citenamefont
  {Kr\"oger},\ and\ \citenamefont {Berndt}}]{ZieglerPRB2009}%
  \BibitemOpen
  \bibfield  {author} {\bibinfo {author} {\bibfnamefont {M.}~\bibnamefont
  {Ziegler}}, \bibinfo {author} {\bibfnamefont {N.}~\bibnamefont {N\'eel}},
  \bibinfo {author} {\bibfnamefont {A.}~\bibnamefont {Sperl}}, \bibinfo
  {author} {\bibfnamefont {J.}~\bibnamefont {Kr\"oger}},\ and\ \bibinfo
  {author} {\bibfnamefont {R.}~\bibnamefont {Berndt}},\ }\bibfield  {title}
  {\bibinfo {title} {Local density of states from constant-current tunneling
  spectra},\ }\href {https://doi.org/10.1103/PhysRevB.80.125402} {\bibfield
  {journal} {\bibinfo  {journal} {Phys. Rev. B}\ }\textbf {\bibinfo {volume}
  {80}},\ \bibinfo {pages} {125402} (\bibinfo {year} {2009})}\BibitemShut
  {NoStop}%
\bibitem [{\citenamefont {H\"ormandinger}(1994)}]{HormandingerPRL1994}%
  \BibitemOpen
  \bibfield  {author} {\bibinfo {author} {\bibfnamefont {G.}~\bibnamefont
  {H\"ormandinger}},\ }\bibfield  {title} {\bibinfo {title} {Comment on
  "{D}irect observation of standing wave formation at surface steps using
  scanning tunneling spectroscopy"},\ }\href
  {https://doi.org/10.1103/PhysRevLett.73.910} {\bibfield  {journal} {\bibinfo
  {journal} {Phys. Rev. Lett.}\ }\textbf {\bibinfo {volume} {73}},\ \bibinfo
  {pages} {910} (\bibinfo {year} {1994})}\BibitemShut {NoStop}%
\bibitem [{\citenamefont {Krenner}\ \emph {et~al.}(2013)\citenamefont
  {Krenner}, \citenamefont {Kühne}, \citenamefont {Klappenberger},\ and\
  \citenamefont {Barth}}]{KrennerSciReports2013}%
  \BibitemOpen
  \bibfield  {author} {\bibinfo {author} {\bibfnamefont {W.}~\bibnamefont
  {Krenner}}, \bibinfo {author} {\bibfnamefont {D.}~\bibnamefont {Kühne}},
  \bibinfo {author} {\bibfnamefont {F.}~\bibnamefont {Klappenberger}},\ and\
  \bibinfo {author} {\bibfnamefont {J.~V.}\ \bibnamefont {Barth}},\ }\bibfield
  {title} {\bibinfo {title} {Assessment of scanning tunneling spectroscopy
  modes inspecting electron confinement in surface-confined supramolecular
  networks},\ }\href {https://doi.org/10.1038/srep01454} {\bibfield  {journal}
  {\bibinfo  {journal} {Scientific Reports}\ }\textbf {\bibinfo {volume} {3}},\
  \bibinfo {pages} {1454} (\bibinfo {year} {2013})}\BibitemShut {NoStop}%
\bibitem [{\citenamefont {Hellenthal}\ \emph {et~al.}(2013)\citenamefont
  {Hellenthal}, \citenamefont {Heimbuch}, \citenamefont {Sotthewes},
  \citenamefont {Kooij},\ and\ \citenamefont {Zandvliet}}]{HellenthalPRB2013}%
  \BibitemOpen
  \bibfield  {author} {\bibinfo {author} {\bibfnamefont {C.}~\bibnamefont
  {Hellenthal}}, \bibinfo {author} {\bibfnamefont {R.}~\bibnamefont
  {Heimbuch}}, \bibinfo {author} {\bibfnamefont {K.}~\bibnamefont {Sotthewes}},
  \bibinfo {author} {\bibfnamefont {E.~S.}\ \bibnamefont {Kooij}},\ and\
  \bibinfo {author} {\bibfnamefont {H.~J.~W.}\ \bibnamefont {Zandvliet}},\
  }\bibfield  {title} {\bibinfo {title} {Determining the local density of
  states in the constant current stm mode},\ }\href
  {https://doi.org/10.1103/PhysRevB.88.035425} {\bibfield  {journal} {\bibinfo
  {journal} {Phys. Rev. B}\ }\textbf {\bibinfo {volume} {88}},\ \bibinfo
  {pages} {035425} (\bibinfo {year} {2013})}\BibitemShut {NoStop}%
\bibitem [{\citenamefont {Pronschinske}\ \emph {et~al.}(2011)\citenamefont
  {Pronschinske}, \citenamefont {Mardit},\ and\ \citenamefont
  {Dougherty}}]{PronschinskePRB2011}%
  \BibitemOpen
  \bibfield  {author} {\bibinfo {author} {\bibfnamefont {A.}~\bibnamefont
  {Pronschinske}}, \bibinfo {author} {\bibfnamefont {D.~J.}\ \bibnamefont
  {Mardit}},\ and\ \bibinfo {author} {\bibfnamefont {D.~B.}\ \bibnamefont
  {Dougherty}},\ }\bibfield  {title} {\bibinfo {title} {Modeling the
  constant-current distance-voltage mode of scanning tunneling spectroscopy},\
  }\href {https://doi.org/10.1103/PhysRevB.84.205427} {\bibfield  {journal}
  {\bibinfo  {journal} {Phys. Rev. B}\ }\textbf {\bibinfo {volume} {84}},\
  \bibinfo {pages} {205427} (\bibinfo {year} {2011})}\BibitemShut {NoStop}%
\bibitem [{\citenamefont {Tersoff}\ and\ \citenamefont
  {Hamann}(1983)}]{TersoffPRL1983}%
  \BibitemOpen
  \bibfield  {author} {\bibinfo {author} {\bibfnamefont {J.}~\bibnamefont
  {Tersoff}}\ and\ \bibinfo {author} {\bibfnamefont {D.~R.}\ \bibnamefont
  {Hamann}},\ }\bibfield  {title} {\bibinfo {title} {Theory and application for
  the scanning tunneling microscope},\ }\href
  {https://doi.org/10.1103/PhysRevLett.50.1998} {\bibfield  {journal} {\bibinfo
   {journal} {Phys. Rev. Lett.}\ }\textbf {\bibinfo {volume} {50}},\ \bibinfo
  {pages} {1998} (\bibinfo {year} {1983})}\BibitemShut {NoStop}%
\bibitem [{\citenamefont {Tersoff}\ and\ \citenamefont
  {Hamann}(1985)}]{TersoffPRB1985}%
  \BibitemOpen
  \bibfield  {author} {\bibinfo {author} {\bibfnamefont {J.}~\bibnamefont
  {Tersoff}}\ and\ \bibinfo {author} {\bibfnamefont {D.~R.}\ \bibnamefont
  {Hamann}},\ }\bibfield  {title} {\bibinfo {title} {Theory of the scanning
  tunneling microscope},\ }\href {https://doi.org/10.1103/PhysRevB.31.805}
  {\bibfield  {journal} {\bibinfo  {journal} {Phys. Rev. B}\ }\textbf {\bibinfo
  {volume} {31}},\ \bibinfo {pages} {805} (\bibinfo {year} {1985})}\BibitemShut
  {NoStop}%
\bibitem [{\citenamefont {Krane}\ \emph {et~al.}(2018)\citenamefont {Krane},
  \citenamefont {Lotze},\ and\ \citenamefont {Franke}}]{KraneSurfSci2018}%
  \BibitemOpen
  \bibfield  {author} {\bibinfo {author} {\bibfnamefont {N.}~\bibnamefont
  {Krane}}, \bibinfo {author} {\bibfnamefont {C.}~\bibnamefont {Lotze}},\ and\
  \bibinfo {author} {\bibfnamefont {K.~J.}\ \bibnamefont {Franke}},\ }\bibfield
   {title} {\bibinfo {title} {Moiré structure of {M}o{S}2 on {A}u(111): Local
  structural and electronic properties},\ }\href
  {https://doi.org/https://doi.org/10.1016/j.susc.2018.03.015} {\bibfield
  {journal} {\bibinfo  {journal} {Surface Science}\ }\textbf {\bibinfo {volume}
  {678}},\ \bibinfo {pages} {136} (\bibinfo {year} {2018})}\BibitemShut
  {NoStop}%
\bibitem [{\citenamefont {Rejali}\ \emph {et~al.}(2022)\citenamefont {Rejali},
  \citenamefont {Farinacci}, \citenamefont {Coffey}, \citenamefont
  {Broekhoven}, \citenamefont {Gobeil}, \citenamefont {Blanter},\ and\
  \citenamefont {Otte}}]{Rejali2022}%
  \BibitemOpen
  \bibfield  {author} {\bibinfo {author} {\bibfnamefont {R.}~\bibnamefont
  {Rejali}}, \bibinfo {author} {\bibfnamefont {L.}~\bibnamefont {Farinacci}},
  \bibinfo {author} {\bibfnamefont {D.}~\bibnamefont {Coffey}}, \bibinfo
  {author} {\bibfnamefont {R.}~\bibnamefont {Broekhoven}}, \bibinfo {author}
  {\bibfnamefont {J.}~\bibnamefont {Gobeil}}, \bibinfo {author} {\bibfnamefont
  {Y.~M.}\ \bibnamefont {Blanter}},\ and\ \bibinfo {author} {\bibfnamefont
  {A.~F.}\ \bibnamefont {Otte}},\ }\bibfield  {title} {\bibinfo {title}
  {Confined vacuum resonances as artificial atoms with tunable lifetime},\
  }\href@noop {} {\bibfield  {journal} {\bibinfo  {journal} {arXiv:2204.10559}\
  } (\bibinfo {year} {2022})}\BibitemShut {NoStop}%
\bibitem [{\citenamefont {Feuchtwang}\ \emph {et~al.}(1983)\citenamefont
  {Feuchtwang}, \citenamefont {Cutler},\ and\ \citenamefont
  {Miskovsky}}]{FeuchtwangPLA1983}%
  \BibitemOpen
  \bibfield  {author} {\bibinfo {author} {\bibfnamefont {T.}~\bibnamefont
  {Feuchtwang}}, \bibinfo {author} {\bibfnamefont {P.}~\bibnamefont {Cutler}},\
  and\ \bibinfo {author} {\bibfnamefont {N.}~\bibnamefont {Miskovsky}},\
  }\bibfield  {title} {\bibinfo {title} {A theory of vacuum tunneling
  microscopy},\ }\href
  {https://doi.org/https://doi.org/10.1016/0375-9601(83)90969-6} {\bibfield
  {journal} {\bibinfo  {journal} {Physics Letters A}\ }\textbf {\bibinfo
  {volume} {99}},\ \bibinfo {pages} {167} (\bibinfo {year} {1983})}\BibitemShut
  {NoStop}%
\bibitem [{\citenamefont {Simmons}(1963)}]{SimmonsJAP1963}%
  \BibitemOpen
  \bibfield  {author} {\bibinfo {author} {\bibfnamefont {J.~G.}\ \bibnamefont
  {Simmons}},\ }\bibfield  {title} {\bibinfo {title} {Generalized formula for
  the electric tunnel effect between similar electrodes separated by a thin
  insulating film},\ }\bibfield  {journal} {\bibinfo  {journal} {J. Appl.
  Phys.}\ }\textbf {\bibinfo {volume} {34}},\ \href
  {https://doi.org/10.1063/1.1702682} {10.1063/1.1702682} (\bibinfo {year}
  {1963})\BibitemShut {NoStop}%
\bibitem [{\citenamefont {Landau}\ and\ \citenamefont
  {Lifshitz}(1977)}]{LandauLifshitz}%
  \BibitemOpen
  \bibfield  {author} {\bibinfo {author} {\bibfnamefont {L.}~\bibnamefont
  {Landau}}\ and\ \bibinfo {author} {\bibfnamefont {E.}~\bibnamefont
  {Lifshitz}},\ }\bibfield  {title} {\bibinfo {title} {Chapter vii - the
  quasi-classical case},\ }in\ \href
  {https://doi.org/https://doi.org/10.1016/B978-0-08-020940-1.50014-1} {\emph
  {\bibinfo {booktitle} {Quantum Mechanics (Third Edition)}}},\ \bibinfo
  {editor} {edited by\ \bibinfo {editor} {\bibfnamefont {L.}~\bibnamefont
  {Landau}}\ and\ \bibinfo {editor} {\bibfnamefont {E.}~\bibnamefont
  {Lifshitz}}}\ (\bibinfo  {publisher} {Pergamon},\ \bibinfo {year} {1977})\
  \bibinfo {edition} {third edition}\ ed.,\ pp.\ \bibinfo {pages}
  {164--196}\BibitemShut {NoStop}%
\bibitem [{\citenamefont {Fowler}\ and\ \citenamefont
  {Nordheim}(1928)}]{NordheimFowler1928}%
  \BibitemOpen
  \bibfield  {author} {\bibinfo {author} {\bibfnamefont {R.~H.}\ \bibnamefont
  {Fowler}}\ and\ \bibinfo {author} {\bibfnamefont {L.}~\bibnamefont
  {Nordheim}},\ }\bibfield  {title} {\bibinfo {title} {Electron emission in
  intense electric fields},\ }\href {https://doi.org/10.1098/rspa.1928.0091}
  {\bibfield  {journal} {\bibinfo  {journal} {Proceedings of the Royal Society
  A}\ }\textbf {\bibinfo {volume} {119}},\ \bibinfo {pages} {173–181}
  (\bibinfo {year} {1928})}\BibitemShut {NoStop}%
\bibitem [{\citenamefont {Ruffieux}\ \emph {et~al.}(2009)\citenamefont
  {Ruffieux}, \citenamefont {Ait-Mansour}, \citenamefont {Bendounan},
  \citenamefont {Fasel}, \citenamefont {Patthey}, \citenamefont {Gr\"{o}ning},\
  and\ \citenamefont {Gr\"{o}ning}}]{RuffieuxPRL2009}%
  \BibitemOpen
  \bibfield  {author} {\bibinfo {author} {\bibfnamefont {P.}~\bibnamefont
  {Ruffieux}}, \bibinfo {author} {\bibfnamefont {K.}~\bibnamefont
  {Ait-Mansour}}, \bibinfo {author} {\bibfnamefont {A.}~\bibnamefont
  {Bendounan}}, \bibinfo {author} {\bibfnamefont {R.}~\bibnamefont {Fasel}},
  \bibinfo {author} {\bibfnamefont {L.}~\bibnamefont {Patthey}}, \bibinfo
  {author} {\bibfnamefont {P.}~\bibnamefont {Gr\"{o}ning}},\ and\ \bibinfo
  {author} {\bibfnamefont {O.}~\bibnamefont {Gr\"{o}ning}},\ }\bibfield
  {title} {\bibinfo {title} {Mapping the electronic surface potential of
  nanostructures surfaces},\ }\href
  {https://doi.org/10.1103/PhysRevLett.102.086807} {\bibfield  {journal}
  {\bibinfo  {journal} {Phys. Rev. Lett.}\ }\textbf {\bibinfo {volume} {102}},\
  \bibinfo {pages} {086807} (\bibinfo {year} {2009})}\BibitemShut {NoStop}%
\bibitem [{\citenamefont {Jung}\ \emph {et~al.}(1995)\citenamefont {Jung},
  \citenamefont {Mo},\ and\ \citenamefont {Himpsel}}]{JungPRL1995}%
  \BibitemOpen
  \bibfield  {author} {\bibinfo {author} {\bibfnamefont {T.}~\bibnamefont
  {Jung}}, \bibinfo {author} {\bibfnamefont {Y.~W.}\ \bibnamefont {Mo}},\ and\
  \bibinfo {author} {\bibfnamefont {F.~J.}\ \bibnamefont {Himpsel}},\
  }\bibfield  {title} {\bibinfo {title} {Identification of metals in scanning
  tunneling microscopy via image states},\ }\href
  {https://doi.org/10.1103/PhysRevLett.74.1641} {\bibfield  {journal} {\bibinfo
   {journal} {Phys. Rev. Lett.}\ }\textbf {\bibinfo {volume} {74}},\ \bibinfo
  {pages} {1641} (\bibinfo {year} {1995})}\BibitemShut {NoStop}%
\bibitem [{\citenamefont {Pivetta}\ \emph {et~al.}(2005)\citenamefont
  {Pivetta}, \citenamefont {Patthey}, \citenamefont {Stengel}, \citenamefont
  {Baldereschi},\ and\ \citenamefont {Schneider}}]{PivettaPRB2005}%
  \BibitemOpen
  \bibfield  {author} {\bibinfo {author} {\bibfnamefont {M.}~\bibnamefont
  {Pivetta}}, \bibinfo {author} {\bibfnamefont {F.}~\bibnamefont {Patthey}},
  \bibinfo {author} {\bibfnamefont {M.}~\bibnamefont {Stengel}}, \bibinfo
  {author} {\bibfnamefont {A.}~\bibnamefont {Baldereschi}},\ and\ \bibinfo
  {author} {\bibfnamefont {W.-D.}\ \bibnamefont {Schneider}},\ }\bibfield
  {title} {\bibinfo {title} {Local work function {M}oir\'e pattern on ultrathin
  ionic films: {N}a{C}l on {A}g(100)},\ }\href
  {https://doi.org/10.1103/PhysRevB.72.115404} {\bibfield  {journal} {\bibinfo
  {journal} {Phys. Rev. B}\ }\textbf {\bibinfo {volume} {72}},\ \bibinfo
  {pages} {115404} (\bibinfo {year} {2005})}\BibitemShut {NoStop}%
\bibitem [{\citenamefont {Ploigt}\ \emph {et~al.}(2007)\citenamefont {Ploigt},
  \citenamefont {Brun}, \citenamefont {Pivetta}, \citenamefont {Patthey},\ and\
  \citenamefont {Schneider}}]{PloigtPRB2007}%
  \BibitemOpen
  \bibfield  {author} {\bibinfo {author} {\bibfnamefont {H.-C.}\ \bibnamefont
  {Ploigt}}, \bibinfo {author} {\bibfnamefont {C.}~\bibnamefont {Brun}},
  \bibinfo {author} {\bibfnamefont {M.}~\bibnamefont {Pivetta}}, \bibinfo
  {author} {\bibfnamefont {F.}~\bibnamefont {Patthey}},\ and\ \bibinfo {author}
  {\bibfnamefont {W.-D.}\ \bibnamefont {Schneider}},\ }\bibfield  {title}
  {\bibinfo {title} {Local work function changes determined by field emission
  resonances: {N}a{C}l/{Ag}(100)},\ }\href@noop {} {\bibfield  {journal}
  {\bibinfo  {journal} {Phys. Rev. B}\ }\textbf {\bibinfo {volume} {76}},\
  \bibinfo {pages} {195404} (\bibinfo {year} {2007})}\BibitemShut {NoStop}%
\bibitem [{\citenamefont {Kubby}\ \emph {et~al.}(1990)\citenamefont {Kubby},
  \citenamefont {Wang},\ and\ \citenamefont {Greene}}]{KubbyPRL1990}%
  \BibitemOpen
  \bibfield  {author} {\bibinfo {author} {\bibfnamefont {J.~A.}\ \bibnamefont
  {Kubby}}, \bibinfo {author} {\bibfnamefont {Y.~R.}\ \bibnamefont {Wang}},\
  and\ \bibinfo {author} {\bibfnamefont {W.~J.}\ \bibnamefont {Greene}},\
  }\bibfield  {title} {\bibinfo {title} {Electron interferometry at a
  heterojunction interface},\ }\href
  {https://doi.org/10.1103/PhysRevLett.65.2165} {\bibfield  {journal} {\bibinfo
   {journal} {Phys. Rev. Lett.}\ }\textbf {\bibinfo {volume} {65}},\ \bibinfo
  {pages} {2165} (\bibinfo {year} {1990})}\BibitemShut {NoStop}%
\bibitem [{\citenamefont {Bobrov}\ \emph {et~al.}(2001)\citenamefont {Bobrov},
  \citenamefont {Mayne},\ and\ \citenamefont {Dujardin}}]{BobrovNature2001}%
  \BibitemOpen
  \bibfield  {author} {\bibinfo {author} {\bibfnamefont {K.}~\bibnamefont
  {Bobrov}}, \bibinfo {author} {\bibfnamefont {A.~J.}\ \bibnamefont {Mayne}},\
  and\ \bibinfo {author} {\bibfnamefont {G.}~\bibnamefont {Dujardin}},\
  }\bibfield  {title} {\bibinfo {title} {Atomic-scale imaging of insulating
  diamond through resonant electron injection},\ }\href
  {https://doi.org/10.1038/35098053} {\bibfield  {journal} {\bibinfo  {journal}
  {Nature}\ }\textbf {\bibinfo {volume} {413}},\ \bibinfo {pages} {616}
  (\bibinfo {year} {2001})}\BibitemShut {NoStop}%
\bibitem [{\citenamefont {Schlenhoff}\ \emph {et~al.}(2020)\citenamefont
  {Schlenhoff}, \citenamefont {Kovarik}, \citenamefont {Krause},\ and\
  \citenamefont {Wiesendanger}}]{SchlenhoffAPL2020}%
  \BibitemOpen
  \bibfield  {author} {\bibinfo {author} {\bibfnamefont {A.}~\bibnamefont
  {Schlenhoff}}, \bibinfo {author} {\bibfnamefont {S.}~\bibnamefont {Kovarik}},
  \bibinfo {author} {\bibfnamefont {S.}~\bibnamefont {Krause}},\ and\ \bibinfo
  {author} {\bibfnamefont {R.}~\bibnamefont {Wiesendanger}},\ }\bibfield
  {title} {\bibinfo {title} {Real-space imaging of atomic-scale spin textures
  at nanometer distances},\ }\href {https://doi.org/10.1063/1.5145363}
  {\bibfield  {journal} {\bibinfo  {journal} {Applied Physics Letters}\
  }\textbf {\bibinfo {volume} {116}},\ \bibinfo {pages} {122406} (\bibinfo
  {year} {2020})}\BibitemShut {NoStop}%
\bibitem [{\citenamefont {Kalff}\ \emph {et~al.}(2016)\citenamefont {Kalff},
  \citenamefont {Rebergen}, \citenamefont {Fahrenfort}, \citenamefont
  {Girovsky}, \citenamefont {Toskovic}, \citenamefont {Lado}, \citenamefont
  {{Fern{\'a}ndez-Rossier}},\ and\ \citenamefont {Otte}}]{KalffNatureNano2016}%
  \BibitemOpen
  \bibfield  {author} {\bibinfo {author} {\bibfnamefont {F.~E.}\ \bibnamefont
  {Kalff}}, \bibinfo {author} {\bibfnamefont {M.~P.}\ \bibnamefont {Rebergen}},
  \bibinfo {author} {\bibfnamefont {E.}~\bibnamefont {Fahrenfort}}, \bibinfo
  {author} {\bibfnamefont {J.}~\bibnamefont {Girovsky}}, \bibinfo {author}
  {\bibfnamefont {R.}~\bibnamefont {Toskovic}}, \bibinfo {author}
  {\bibfnamefont {J.~L.}\ \bibnamefont {Lado}}, \bibinfo {author}
  {\bibfnamefont {J.}~\bibnamefont {{Fern{\'a}ndez-Rossier}}},\ and\ \bibinfo
  {author} {\bibfnamefont {A.~F.}\ \bibnamefont {Otte}},\ }\bibfield  {title}
  {\bibinfo {title} {A kilobyte rewritable atomic memory},\ }\href
  {https://doi.org/10.1038/nnano.2016.131} {\bibfield  {journal} {\bibinfo
  {journal} {Nature Nanotechnology}\ }\textbf {\bibinfo {volume} {11}},\
  \bibinfo {pages} {926} (\bibinfo {year} {2016})}\BibitemShut {NoStop}%
\bibitem [{\citenamefont {Girovsky}\ \emph {et~al.}(2017)\citenamefont
  {Girovsky}, \citenamefont {Lado}, \citenamefont {Otte}, \citenamefont
  {Kalff}, \citenamefont {Fahrenfort}, \citenamefont {Peters},\ and\
  \citenamefont {{Fern{\'a}ndez-Rossier}}}]{GirovskySciPost2017}%
  \BibitemOpen
  \bibfield  {author} {\bibinfo {author} {\bibfnamefont {J.}~\bibnamefont
  {Girovsky}}, \bibinfo {author} {\bibfnamefont {J.}~\bibnamefont {Lado}},
  \bibinfo {author} {\bibfnamefont {S.}~\bibnamefont {Otte}}, \bibinfo {author}
  {\bibfnamefont {F.}~\bibnamefont {Kalff}}, \bibinfo {author} {\bibfnamefont
  {E.}~\bibnamefont {Fahrenfort}}, \bibinfo {author} {\bibfnamefont
  {L.}~\bibnamefont {Peters}},\ and\ \bibinfo {author} {\bibfnamefont
  {J.}~\bibnamefont {{Fern{\'a}ndez-Rossier}}},\ }\bibfield  {title} {\bibinfo
  {title} {Emergence of quasiparticle {{Bloch}} states in artificial crystals
  crafted atom-by-atom},\ }\href {https://doi.org/10.21468/SciPostPhys.2.3.020}
  {\bibfield  {journal} {\bibinfo  {journal} {SciPost Physics}\ }\textbf
  {\bibinfo {volume} {2}},\ \bibinfo {pages} {020} (\bibinfo {year}
  {2017})}\BibitemShut {NoStop}%
\bibitem [{\citenamefont {Drost}\ \emph {et~al.}(2017)\citenamefont {Drost},
  \citenamefont {Ojanen}, \citenamefont {Harju},\ and\ \citenamefont
  {Liljeroth}}]{DrostNaturePhys2017}%
  \BibitemOpen
  \bibfield  {author} {\bibinfo {author} {\bibfnamefont {R.}~\bibnamefont
  {Drost}}, \bibinfo {author} {\bibfnamefont {T.}~\bibnamefont {Ojanen}},
  \bibinfo {author} {\bibfnamefont {A.}~\bibnamefont {Harju}},\ and\ \bibinfo
  {author} {\bibfnamefont {P.}~\bibnamefont {Liljeroth}},\ }\bibfield  {title}
  {\bibinfo {title} {Topological states in engineered atomic lattices},\ }\href
  {https://doi.org/10.1038/nphys4080} {\bibfield  {journal} {\bibinfo
  {journal} {Nature Physics}\ }\textbf {\bibinfo {volume} {13}},\ \bibinfo
  {pages} {668} (\bibinfo {year} {2017})}\BibitemShut {NoStop}%
\bibitem [{\citenamefont {Stepanow}\ \emph {et~al.}(2011)\citenamefont
  {Stepanow}, \citenamefont {Mugarza}, \citenamefont {Ceballos}, \citenamefont
  {Gambardella}, \citenamefont {Aldazabal}, \citenamefont {Borisov},\ and\
  \citenamefont {Arnau}}]{StepanowPRB2011}%
  \BibitemOpen
  \bibfield  {author} {\bibinfo {author} {\bibfnamefont {S.}~\bibnamefont
  {Stepanow}}, \bibinfo {author} {\bibfnamefont {A.}~\bibnamefont {Mugarza}},
  \bibinfo {author} {\bibfnamefont {G.}~\bibnamefont {Ceballos}}, \bibinfo
  {author} {\bibfnamefont {P.}~\bibnamefont {Gambardella}}, \bibinfo {author}
  {\bibfnamefont {I.}~\bibnamefont {Aldazabal}}, \bibinfo {author}
  {\bibfnamefont {A.~G.}\ \bibnamefont {Borisov}},\ and\ \bibinfo {author}
  {\bibfnamefont {A.}~\bibnamefont {Arnau}},\ }\bibfield  {title} {\bibinfo
  {title} {Localization, splitting, and mixing of field emission resonances
  induced by alkali metal clusters on {C}u(100)},\ }\href@noop {} {\bibfield
  {journal} {\bibinfo  {journal} {Phys. Rev. B}\ }\textbf {\bibinfo {volume}
  {83}},\ \bibinfo {pages} {115101} (\bibinfo {year} {2011})}\BibitemShut
  {NoStop}%
\bibitem [{\citenamefont {Crampin}(2005)}]{CrampinPRL2005_2}%
  \BibitemOpen
  \bibfield  {author} {\bibinfo {author} {\bibfnamefont {S.}~\bibnamefont
  {Crampin}},\ }\bibfield  {title} {\bibinfo {title} {Lifetimes of
  stark-shifted image states},\ }\href
  {https://doi.org/10.1103/PhysRevLett.95.046801} {\bibfield  {journal}
  {\bibinfo  {journal} {Phys. Rev. Lett.}\ }\textbf {\bibinfo {volume} {95}},\
  \bibinfo {pages} {046801} (\bibinfo {year} {2005})}\BibitemShut {NoStop}%
\bibitem [{\citenamefont {Berghaus}\ \emph {et~al.}(1988)\citenamefont
  {Berghaus}, \citenamefont {Brodde}, \citenamefont {Neddermeyer},\ and\
  \citenamefont {Tosch}}]{BerghausSurfSci1988}%
  \BibitemOpen
  \bibfield  {author} {\bibinfo {author} {\bibfnamefont {T.}~\bibnamefont
  {Berghaus}}, \bibinfo {author} {\bibfnamefont {A.}~\bibnamefont {Brodde}},
  \bibinfo {author} {\bibfnamefont {H.}~\bibnamefont {Neddermeyer}},\ and\
  \bibinfo {author} {\bibfnamefont {S.}~\bibnamefont {Tosch}},\ }\bibfield
  {title} {\bibinfo {title} {Scanning tunneling microscopy and spectroscopy on
  7 × 7 reconstructed si(111) surfaces containing defects},\ }\href
  {https://doi.org/https://doi.org/10.1016/0039-6028(88)90334-2} {\bibfield
  {journal} {\bibinfo  {journal} {Surface Science}\ }\textbf {\bibinfo {volume}
  {193}},\ \bibinfo {pages} {235} (\bibinfo {year} {1988})}\BibitemShut
  {NoStop}%
\bibitem [{\citenamefont {Stroscio}\ and\ \citenamefont
  {Kaiser}(1993)}]{StroscioKaiser1993}%
  \BibitemOpen
  \bibinfo {editor} {\bibfnamefont {J.~A.}\ \bibnamefont {Stroscio}}\ and\
  \bibinfo {editor} {\bibfnamefont {W.~J.}\ \bibnamefont {Kaiser}},\ eds.,\
  \href {https://doi.org/https://doi.org/10.1016/S0076-695X(08)60001-0} {\emph
  {\bibinfo {title} {Scanning Tunneling Microscopy}}},\ \bibinfo {series}
  {Methods in Experimental Physics}, Vol.~\bibinfo {volume} {27}\ (\bibinfo
  {publisher} {Academic Press},\ \bibinfo {year} {1993})\BibitemShut {NoStop}%
\end{thebibliography}%

\end{document}